\documentclass[12pt]{article}

\usepackage[utf8]{inputenc}
\usepackage[margin=1.0in]{geometry}
\usepackage{authblk}
\usepackage{soul}
\usepackage{multicol}
\usepackage{graphicx}
\usepackage{subfigure}
\usepackage{wrapfig, blindtext}
\usepackage{lineno}
\usepackage{xcolor}
\usepackage{float}

\usepackage{caption}
\def\xmm {{\it XMM-Newton}~}

\def\suzaku {{\it Suzaku}~}
\def\suzakun {{\it Suzaku}}

\def\erosita {{\it eROSITA}~}

\def\fermi {{\it Fermi}~}

\def\eb {{\it eROSITA~bubbles}~}
\def\ebn {{\it eROSITA~bubbles}}
\def\fb {{\it Fermi~bubbles}~}
\def\fbn {{\it Fermi~bubbles}}

\def\oi{{{\rm O}\,{\sc i}~}}

\def\ovii{{{\rm O}\,{\sc vii}~}}

\def\oviii{{{\rm O}\,{\sc viii}~}}

\def\neix{{{\rm Ne}\,{\sc ix}~}}

\def\nvii{{{\rm N}\,{\sc vii}~}}

\title{Thermal and chemical properties of the eROSITA bubbles from Suzaku observations}
\author[1, 2]{Anjali~Gupta \thanks{agupta1@cscc.edu}}
\author[2, 3]{Smita~Mathur}
\author[1, 2]{Joshua~Kingsbury}
\author[2]{Sanskriti~Das}
\author[4]{Yair~Krongold}

\affil[1]{Columbus State Community College, 550 E Spring St., Columbus, OH 43215, USA}

\affil[2]{Department of Astronomy, The Ohio State University, 140 West 18th Avenue, Columbus, OH 43210, USA} 

\affil[3]{Center for Cosmology and Astro-Particle Physics, The Ohio State University, 140 West 18th Avenue, Columbus, OH 43210, USA} 

\affil[4]{Instituto de Astronomia, Universidad Nacional Autonoma de Mexico, 04510 Mexico City, Mexico} 

\begin{document}

\maketitle

\noindent
{\bf
The X-ray bright bubbles at the Galactic Center provide an opportunity to understand the effects of feedback on galaxy evolution. The shells of the eROSITA bubbles show enhanced X-ray emission over the sky background. Previously, these shells were assumed to have a single temperature component and to trace the shock-heated lower-temperature halo gas. Using Suzaku observations,  we show that the X-ray emission of the shells is more complex and best described by a two-temperature thermal model: one component close to the Galaxy’s virial temperature and the other at  super-virial temperatures. Furthermore, we demonstrate that temperatures of the virial and super-virial components are similar in the shells and in the ambient medium, although the emission measures are significantly higher in the shells. This leads us to conclude that the eROSITA bubble shells are X-ray bright because they trace denser gas, not because they are hotter. Given that the pre- and post-shock temperatures are similar and the compression ratio of the shock is high, we rule out that the bubble shells trace adiabatic shocks, in contrast to what was assumed in previous studies. We also observe non-solar Ne/O and Mg/O ratios in the shells, favouring stellar feedback models for the formation of the bubbles and settling a long-standing debate on their origin.
}
\newpage

\vspace{0.5cm}
\noindent
{\Large{\bf Introduction}}\\
The all-sky survey performed by the \erosita X-ray telescope has revealed a large hourglass-shaped structure in the center of the  Milky-way (MW; \cite{Predehl:2020}), called the \ebn. The X-ray bright quasi-circular feature in the northern sky, which include structures such as the North Polar Spur and the Loop-I, have been known since their discovery by ROSAT \cite{Snowden:1995}. The \erosita map shows X-ray emission from a similarly huge quasi-circular annular structure in the southern sky; together they seem to form giant Galactic X-ray bubbles emerging from the Galactic Center (GC). 

The large-scale X-ray emission observed by \erosita in its medium energy band ($\rm 0.6-1.0 ~keV$) shows that the intrinsic size of the bubbles is several kiloparsecs across \cite{Predehl:2020}. The \eb show striking morphological similarities to the well-known \fb detected in $\gamma-$ray by the \fermi telescope \cite{Su:2010}, but they are larger and more energetic. The \fermi and \erosita bubbles (collectively we called the `Galactic bubbles') provide an exciting laboratory for studying the feedback because of their size and the location in the Galaxy. These bubbles are magnificent structures injecting energy/momentum into the MW circum-galactic medium (CGM) or halo. (The CGM of the Milky Way is usually referred as the Galactic ``halo''.  CGM is a more prevalent term for external galaxies.  Both the terms mean essentially the same, and we will use these terms interchangeably). In order to understand the feedback process, it is important to determine the thermal, kinetic, and dynamical structure of these bubbles. 

The Galactic bubbles are expanding into the MW halo; we therefore examine the spatial distribution of the X-ray emission from the bubble shells and from the halo around them to constrain their thermal structure.  We conducted a survey of \suzaku observations with this goal. We selected $230$ archival \suzaku observations of the soft diffuse X-ray background (SDXB)  to characterize the X-ray emission from the  Galactic bubbles (Galactic longitudes $300^{\circ} < l < 60^{\circ}$) and from the surrounding extended halo ($60^{\circ} < l < 300^{\circ}$). 

In order to extract the Galactic bubbles/halo emission from the SDXB, it is crucial to accurately model the other components of the SDXB, such as the Local Bubble (LB), solar wind charge exchange (SWCX), the cosmic X-ray background (CXB), and the instrumental background. We included emission from these components in the spectral fitting (see methods). 

\vspace{0.5cm}
\noindent
{\Large{\bf Results}}\\

\noindent
{\bf A Two-Temperature Spectral Model}\\
\noindent
Typically the Galactic bubbles/halo emission is described by a single temperature thermal component. However, our spectral fits to the \suzaku spectra show that the X-ray emission of the bubble shells as well as the outer halo is best described by two-thermal components (see methods), a warm-hot phase near the Galaxy's virial temperature $\rm kT \approx 0.2 ~keV ~(2.3 \times 10^{6}~K)$ and a hot phase at super-virial temperatures ranging between $\rm kT = 0.4-1.1~keV ~(0.5-1.3 \times 10^{7}~K)$. Fig. 1 shows the X-ray emission maps of the warm-hot and the hot components of the Galactic bubbles and the surrounding halo emission.

Out of our $\rm 150$ sightlines that probe the Galactic bubbles region, the  hot thermal component is required at high confidence (F-test probability) of $>99.99\%$ in 55 sightlines, at $>90.0\%$ in 80 sightlines, and  at $1\sigma$ significance in 127 sightlines. For the regions outside the bubbles, the hot component is required  at the confidence of $>99.0\%$ in 26 sightlines, at $>90.0\%$ in 51 sightlines, and at $1\sigma$ significance in 64 sightlines, out of our $80$ sightlines. Fig. 2 shows the hot-component F-test probability map for all the sightlines investigated in this work. 

The best fit models also require overabundance of nitrogen at the confidence of $>90.0\%$ (average $\rm N = 4.2\pm0.2~ solar$) in the warm-hot phase, both from and around the bubbles. Toward the 10 Galactic bubbles sightlines (but not outside the bubbles), the best fit model also requires super-solar abundance ratios of neon to oxygen at $>1\sigma$ significance (average $\rm Ne = 2.1\pm0.2~ solar$).  Supersolar magnesium to oxygen ratio ($\rm Mg = 3.6\pm1.4~ solar$) is also required along 1 sightline.

The presence of the warm-hot, virial-temperature gas in the Galactic halo has been known for years \cite{Nicastro:2002, Williams:2005, Williams:2006, Williams:2007, Gupta:2012, Mathur:2023}, however the super-virial temperature gas was recently discovered. The first robust detection was in the sightline to $1ES1553+113$ passing close to the North Polar Spur/Loop-I region of the Galactic bubbles \cite{Das:2019a, Das:2019b}. Later, the similar temperature hot gas was detected toward three other sightlines passing close to and away from the Galactic bubbles \cite{Gupta:2021}. These studies showed the presence of the hot gas in the Galactic halo, but it was not known how ubiquitous it is.

In this work we have detected the hot gas toward a large number of sightlines distributed all over the sky. We confirmed with high confidence that the super-virial temperature plasma is widespread in the Galaxy and it is not necessarily associated with the Galactic bubbles only (Fig. 2). This has significant implications for our understanding of the bubbles. 

The Galactic bubbles are believed to have formed by the GC feedback (e.g., \cite{Sofue:2000, Crocker:2015, Sarkar:2016}), that has generated shocks in the northern and the southern hemispheres, and these shocks have been expanding into the Galactic halo. The shape and speed of shocks travelling through the MW CGM depend on the CGM density, pressure and temperature. Thus to characterize the properties of the shocks, we examined the variation in thermal parameters of the warm-hot and the hot phases of the shocked (bubble shells) and unshocked (outer halo) plasma of the Galactic halo. 

Fig. 3 shows the distribution of emission measures (EMs) and temperatures of both the thermal components as a function of Galactic longitude. We see that the EMs are significantly higher for sightlines piercing the bubbles compared to the outer halo sightlines. However, the temperatures of the warm-hot and hot components are similar in/outside the shells. X-ray surface brightness of a gaseous medium depends on its temperature as well as the EM. Our results show that the Galactic bubble shells have higher EMs but not higher temperature in comparison to the surrounding halo, contrary to the current proposed models of the bubbles \cite{Predehl:2020}. Since the EM is proportional to the density square, we argue that the higher X-ray surface brightness of the Galactic bubble shells as seen in the \erosita all-sky map is a result of the compressed denser gas, but it is not hotter than the surrounding medium. 

\vspace{0.5cm}
\noindent
{\bf The shock properties}\\
\noindent
The \eb are likely produced by shocks that have been driven into the northern and the southern Galactic halo. The speed and shape of shocks depend on the total energy input and the thermal parameters of the ambient plasma. Multiple studies have attempted to characterize the X-ray emission from the Galactic bubbles \cite{Predehl:2020, Kataoka:2013, Kataoka:2015, Kataoka:2018, Kataoka:2021, Tahara:2015, Akita:2018}. These authors assumed a single temperature for the X-ray emitting shells, and measured it to be $\rm \sim 0.3~keV$. They interpreted that this emission arises in the weakly shock-heated Galactic halo gas at $\rm T \sim 0.2~keV$, and they estimated a Mach number of the shock of $\rm M \approx 1.5$ using the Rankine–Hugoniot (R-H) conditions for the assumed temperature \cite{Predehl:2020, Kataoka:2013, Kataoka:2021}. 

We have found that the X-ray spectral model is more complex than previously assumed. The shells are best described by a two temperature model and the temperatures in and around the shells is similar. This shows that the shells are not shock {\it heated}; the shells are bright because they trace denser gas, not hotter gas. We compared the thermal parameters of the bubble shells and the pre-shock halo gas to infer the shock properties further (following \cite{Draine:2011}). The gas density of the \eb shells are estimated from the measured values of the EMs. The EM is given by $n^{2}L$, where $n$ is the density (assuming a uniform medium) and $L$ is the line of sight pathlength. The average line of sight path-length is about $\rm L \sim 5~kpc$, for a shell of outer radius of $\rm \sim 7~kpc$ and thickness of $\rm \sim 4~kpc$ (from \cite{Predehl:2020}). This results in an average density of $\rm n_{shell} \sim 1.6 \times 10^{-3}~cm^{-3}$ within the shells. The Galactic halo studies (both observational and theoretical) have estimated the halo density of about $\rm 2-5 \times 10^{-4}~cm^{-3}$ at a distance of $\rm 10 ~kpc$ (approximate location of the shells) from the GC \cite{Gupta:2012, Fang:2013, Faerman:2017, Faerman:2020}. Adopting the unperturbed halo density of $\rm n_{o}=4 \times 10^{-4}~cm^{-3}$ (the same as used by \cite{Predehl:2020}), we calculated the compression ratio of shock of $\rm \sim 4.0$.
For a weak adiabatic shock, the post-shock density can only marginally increase according to the Rankine–Hugoniot (R-H) condition for density. The large compression ratio we measure is inconsistent with the assumption of a weak adiabatic shock in \cite{Predehl:2020}. 

Furthermore, the estimated $\rm 0.3~keV$ plasma density of $\rm 0.002~cm^{-3}$ in \cite{Predehl:2020} is a factor of about $\rm 5$ times larger than their adopted value of the pre-shocked halo density of $\rm 4 \times 10^{-4}~cm^{-3}$. However, according to the R-H condition for density for a non-radiative shock of $\rm M=1.5$, the density ratio should be $\sim 1.7$ instead. Even in the limit of a very strong shock $\rm M \rightarrow \infty$, the density jump for a non-radiative shock is bounded by a value of $(\gamma +1 )/(\gamma -1)$ which equals 4 for $\gamma = 5/3$, and cannot be as high as $\rm 5$. Thus we see that the shocks cannot be adiabatic. 

Detailed theoretical calculations of the shock properties of the \eb are beyond the scope of this paper. Any successful model of these enigmatic bubbles must explain the observed thermal and chemical properties presented in this paper.

\vspace{0.5cm}
\noindent
{\Large{\bf Discussion}}\\

\noindent
{\bf Comparison with Previous Studies}\\
Previous studies used a single temperature model with fixed relative abundances to define the X-ray emission and inferred that the Galactic bubble shells have temperature of $\rm kT \approx 0.3~keV$ \cite{Kataoka:2013, Kataoka:2015, Kataoka:2018, Kataoka:2021, Tahara:2015, Akita:2018} or $\rm kT \approx 0.4~keV$ \cite{Miller:2016}. This is higher than the temperature of the MW CGM of $\rm \sim 0.2~keV$, which led them to conclude that the bubble shells represent shock heated gas. Further, using the ratio of the pre- and post-shock temperatures, these works estimated the shock speed, age and energy of the bubbles. We show that the use of a single temperature model to represent the shell emission was too simplistic, leading to incorrect physical model of the bubbles. 
In this work using the better spectral models we accurately measured the temperatures, EMs and relative metal abundances of the plasma in the bubble shells. 

We estimated that the average density of the warm-hot component of bubbles is about 4 times larger than the pre-shock halo gas. For an adiabatic (non-radiative) shock, the maximum density jump possible is equal to 4, in the limit of a very strong shock with Mach number $\rm M \rightarrow \infty$. But such a strong shock should also cause a significant increase in temperature. Given that the pre- and post-shock temperatures are similar, and the compression ratio of the shock is high, we rule out that the bubble shells trace adiabatic shocks, in contrast to what was assumed in \cite{Predehl:2020} (see methods). 

\vspace{0.5cm}
\noindent
{\bf AGN or Stellar feedback?}\\
\noindent
The physical origin of the Galactic bubbles is still under debate. Since the discovery of the \fbn, there have been a lot of efforts to understand the formation mechanism of the bubbles, with several theoretical models proposed in literature. On the basis of their feedback mechanisms, these models can be broadly divided into two categories; one is the nuclear star-forming activity similar to starburst galaxies and the other is the past AGN activity of the GC supermassive black hole. 

Metal abundance measurements provide a useful insight on the origin of the bubbles. In the star-formation activity scenarios, the bubbles are enriched by metals produced by SNe and stellar winds, whose abundances are different from that in the interstellar medium (ISM). On the other hand, in the AGN wind scenario, the abundance of the wind would be the same as the ambient ISM which accretes onto the GC supermassive black hole. In this work, we have measured super-solar abundances of neon and magnesium compared to oxygen toward a few sightlines passing through the bubbles; this supports the star-formation related feedback scenario for the formation of the Galactic bubbles. 

\newpage
\noindent
{\Huge {\bf Methods}}\\

\noindent
{\Large{\bf Data Selection and Reduction}}\\

\noindent
In this work we analyzed the \suzaku archival observations probing the eROSITA bubbles regions towards the center of the Galaxy, as well as the surrounding fields. For the Galactic bubbles regions we selected observations with exposure time of $\rm \geq 20~ks$. As can be seen in the \erosita all sky map, the surrounding fields are much fainter in X-rays; therefore we selected the observations with higher exposure time of $\rm \geq 50~ks$. Further, to avoid the contamination from the Galactic disk, we chose targets at least 15 degrees above/below the Galactic plane. This yielded multiple observations of $150$ and $80$ fields, probing the Galactic bubbles and the surrounding regions, respectively. 

We performed \suzaku data reduction with HEAsoft version $\rm 6.29$. We only used the data from the back-illuminated (BI) X-ray imaging spectrometer~1 (XIS1) detector, as this has better sensitivity at low energies than the front-illuminated (FI) XIS0 and XIS3 detectors. We combined the data taken in the $\rm 3 \times 3$ and $\rm5 \times 5$  observation modes. We applied extra screening to the data in addition to the standard screening described in the Suzaku Data Reduction Guide. To minimize the detector background, we excluded times when the cut-of-rigidity (COR) of the Earth's magnetic field was less than 8~GV (the default value is $\rm COR = 2 ~GV$). Further, we increased the filter value of the angle between \suzaku's line-of-sight and the limb of the Earth (ELV), from the default value of $\rm 5^{\circ}$ to $\rm 10^{\circ}$. This minimizes the excess events in the $\rm 0.5-0.6~keV$ band due to solar X-rays scattered off the Earth's atmosphere \cite{Smith:2007}.

The activity of our own Sun can affect the space weather and contaminate data taken by space observatories. The Sun was at its minimum in the 11 year solar activity cycle when \suzaku was launched on July 10, 2005, approaching its maximum from early 2011 to 2014. Solar X-rays interact with the neutral oxygen in the earth’s atmosphere and generate a fluorescent emission line at $\rm 0.525~keV$ \cite{Sekiya:2014}. This line in the soft-X-ray band can be detected by instruments on board satellites in the low-Earth orbits, like \suzakun. Gupta et al. \cite{Gupta:2021} reported that in four \suzaku spectra taken in 2014, the \oi intensity was about 25\% to 130\% of the \ovii intensity (at the temperatures of a few million kelvin, the \ovii and \oviii emission lines are the dominant features characterizing the MW CGM or the bubbles). The \oi contamination can be minimized by removing events taken during time intervals when the elevation angle from the bright Earth limb (the DYE\_ELV parameter) is larger \cite{Sekiya:2014}, as we did. 

For observations taken in 2011-2015, we carefully quantified the \oi fluorescence line contamination in our analysis (for details see \cite{Gupta:2021}). We examined the \oi emission with respect to different DYE\_ELV values ($\rm > 20^{\circ}$, $\rm > 40^{\circ}$, and $\rm > 60^{\circ}$) and selected the best value for the DYE\_ELV parameter that provided a good balance between optimizing the effective exposure time and mitigating the \oi contamination. We then model the residual \oi emission with a gaussian line in the spectral analysis. For observations taken before 2011 we applied standard screening of $\rm DYE\_ELV > 20^{\circ}$. 

The goal of this work is to analyze the diffuse emission, hence it is important to remove point sources. We generated the $\rm 0.5-2.0~keV$ images and identified the bright point sources. We selected the point source exclusion regions of radii  of $\rm 1'-3'$ (c.f \suzaku XRT's half-power diameter of  1.8' to 2.3').  Then we extracted the diffuse emission spectrum from the entire field-of-view after excluding the point source regions. 
We produced the redistribution matrix files (RMFs) using the {\it xisrmfgen} ftool, in which the degradation of energy resolution and its position dependence are included. We also prepared ancillary response files (ARFs) using {\it xissimarfgen} ftool. For the ARF calculations we assumed a uniform source of radius $\rm 20''~$ and used a detector mask which removed the bad pixel regions. We estimate the total instrumental background from the database of the night Earth data with the {\it xisnxbgen} ftool. 

\vspace{0.5cm}
\noindent
{\Large{\bf Spectral Analysis}}\\

\noindent

We performed all the spectral fitting with Xspec version 12.11.1 \cite{Arnaud:1996}. We modeled all the thermal plasma components in collisional ionization equilibrium (CIE) with the APEC (version 3.0.9) model and used solar relative metal abundances  \cite{Anders:1989}. For absorption by the Galactic disk, we used the {\it phabs} model in XSPEC.

\suzaku provides an opportunity to resolve the different components of the SDXB due to its low and stable detector background even at low energies ($\rm 0.3-1.0~keV$). 
The SDXB spectrum is usually described by a three-component model consisting of: 1) a foreground component of LB and SWCX, modeled as an unabsorbed thermal plasma emission in CIE, 2) a background component of CXB (made of unresolved point sources) modeled with an absorbed power-law, and 3) the MW halo emission, modeled as an equilibrium thermal plasma absorbed by the cold gas in the Galactic disk (the halo emission toward the Galactic center is dominated by the bubbles; \cite{Predehl:2020}). Recently we found that in few observations an additional absorbed thermal component and/or  enhanced Ne abundance is required to explain the excess emission near $\rm 0.7-0.9~keV$  in the \suzaku \cite{Gupta:2021} and  \xmm \cite{Das:2019b} SDXB spectra.

We started with fitting the \suzaku SDXB spectra with a three-component model. The temperature of the foreground component was frozen at $\rm kT=0.1~keV$ (e.g., \cite{ McCammon:2002, Gupta:2009, Henley:2013, Liu:2017}), but we allowed the normalization to vary. We modeled the Galactic bubbles (or the extended CGM) emission as single temperature collisionally ionized plasma characterized by temperature (kT) and emission measure (EM), and with fixed metallicity. The X-ray emission data does not contain any line or edge of hydrogen. Thus we cannot obtain absolute metal abundances from X-ray emission data alone. Instead, the X-ray observations provide constraints on relative metal abundances, for example N/O, C/O, Ne/O. We fixed the total metallicity to 1 (in solar units) for both the thermal components as the total metallicity and normalizations (or EM) are degenerate in the APEC model. We allowed the power-law photon index and the normalization to vary in the spectral fits. 

This three-component model provided a poor fit to most of the data sets, showing strong excess emission at low ($\rm \sim 0.4-0.5 ~keV$) and high ($\rm 0.8-1.0~keV$) energy bands. An example of the \suzaku spectrum for one observation showing these excess emissions is shown in Supplementary Fig. 1 (top panel).  

Since \nvii and \neix have strong transitions at $\rm 0.5~keV$ and $\rm 0.9~keV$, respectively, we allowed  the nitrogen and neon relative abundances to vary in our above model. That provided a slightly better fit but still left significant excess emission at the higher energy side ($\rm 0.8-1.0~keV$). To fit the higher energy excess emission we added an additional thermal component to our model. This significantly improved the fit for most of our data sets. An example of best fit two-temperature model is shown in Supplementary Fig. 1 (bottom panel). The temperature of the second thermal component is much higher ($\rm kT = 0.4-1.1 ~keV$) than the first  ($\rm kT \sim 0.2~keV$ known as the warm-hot component); we call this the hot component. 

A recent study \cite{Yamamoto:2022} has shown that the CGM spectra can be fitted with a non-equilibrium ionization (NEI) model. This, in principle, could be an alternative to our two-temperature model. To test this possibility, we fitted our data with the NEI model, but found that the fits were significantly worse. Therefore, we use our results of the two-temperature model in all the further discussion. 

We have used abundances from \cite{Anders:1989} in the above analysis. We obtained similar results using the abundances from \cite{Lodders:2009}.

\vspace{0.5cm}
\noindent
{\Large{\bf Distribution of Thermal Parameters}}\\

\noindent
{\bf Galactic Bubbles Region:} 
The temperature of the warm-hot component from the bubble shells is consistent within errors with an average value of $\rm kT= 0.205\pm0.003\pm0.002~keV$ (statistical and systematic errors). The EMs of the warm-hot component of the bubbles regions varies greatly in the range  $\rm 2.2-46.9 \times 10^{-3}~cm^{-6}~pc$ with a mean of $\rm 13.9 \times 10^{-3}~cm^{-6}~pc$ (and median of $\rm 12.7 \times 10^{-3}~cm^{-6}~pc$). Overabundance of nitrogen by $1.3-10.3~$ solar in the warm-hot phase is required for most of the observations that are not contaminated by the local \oi emission. In observations contaminated by \oi we were not able to constrain the nitrogen abundance, therefore we fixed that to solar. A few sightlines also require super-solar abundances of neon and magnesium compared to oxygen. 

The measured temperatures and EMs of the hot gas in the bubble regions are in the range of $\rm 0.4-1.1~keV$ and $\rm 0.4-13.9 \times 10^{-3}~cm^{-6}~pc$, respectively, with mean values of $\rm 0.741\pm0.018~keV$ and $\rm 2.3 \times 10^{-3}~cm^{-6}~pc$.  The emission from the hot component is significantly fainter than that from the warm-hot component. 

\vspace{0.15cm}
\noindent
{\bf Extended Halo Region:} The warm-hot component has a uniform temperature of $\rm kT = 0.201\pm0.004\pm0.003~keV$, similar to those in the bubbles regions. The hot component has a temperatures in the range of $\rm kT = 0.4-1.2~keV$, with mean values of $\rm 0.837\pm0.028~keV$. The average temperature of the hot component is slightly lower in the bubbles region in comparison to the outer halo, though the two are consistent with each other within $\rm 3\sigma$. However, for both components, the EMs are much lower in the extended halo regions compared to the EMs from the bubbles regions. The EMs of the warm-hot phase are in the range of $\rm 0.8-14.2 \times 10^{-3}~cm^{-6}~pc$ with a mean of $\rm 4.4 \times 10^{-3}~cm^{-6}~pc$. The hot phase EMs are much lower,  with the range of $\rm 0.2-1.5 \times 10^{-3}~cm^{-6}~pc$ with a mean of $\rm 6.1 \times 10^{-4}~cm^{-6}~pc$. We also found that nitrogen is overabundant by $1.0-11.4~$ solar in the warm-hot phase. However, super-solar abundances of neon and magnesium compared to oxygen are not required toward any of the sightlines.

Thermal parameters of the Galactic bubbles and the extended halo regions are given in the Supplementary Table 1.

\vspace{0.15cm}
\noindent
{\bf Northern vs Southern Bubbles:} We compared the thermal properties of the northern ($b>15^{\circ}$) and the southern ($b<-15^{\circ}$) bubbles. We have plotted the temperatures and EMs of the Galactic bubbles sightlines vs the Galactic latitude in Supplementary Fig. 2. The sightlines probing the northern bubble have comparatively higher EMs than the southern bubble, but their temperatures are similar. 
For the northern bubble, the warm-hot and the hot components have the average temperatures of  $\rm 0.203\pm0.003\pm0.002~keV$ and $\rm 0.734\pm0.018\pm0.010~keV$ and the average EMs of $\rm 14.8\pm0.9\pm0.2 \times 10^{-3}~cm^{-6}~pc$ and $\rm 2.5\pm0.2\pm0.1 \times 10^{-3}~cm^{-6}~pc$, respectively. The southern bubble have similar temperatures of $\rm 0.210\pm0.005\pm0.003~keV$ and $\rm 0.759\pm0.024\pm0.020~keV$, but have lower EMs of $\rm 9.4\pm1.1\pm0.3\times 10^{-3}~cm^{-6}~pc$ and $\rm 1.6\pm0.4\pm0.1 \times 10^{-3}~cm^{-6}~pc$, for the warm-hot and hot phases, respectively.

The EM of the warm-hot component decreases with Galactic latitude out to about {\bf{\it b}}$~\pm 45$, then becomes  comparatively uniform. The hot component EM variation shows the similar trend but is less prominent. The decrease in the EM with Galactic latitude is in agreement with the \erosita X-ray emission all-sky map, which shows very bright emission at the base of the bubbles, with the surface-brightness falling monotonically away from the base. We do not find any such relation in the distribution of temperatures with the Galactic latitude. This further confirms that regions with brighter emission in the \erosita all-sky map have higher EMs but are not hotter than the surrounding medium.

For the northern and the southern bubbles the total X-ray surface brightness ($\rm 0.5-2.0~keV$) of the warm-hot component is $\rm 3.1\pm0.6 \times 10^{-15}~ergs~cm^{-2}~s^{-1}~arcmin^{-2}$ . Assuming the projected area of the \eb of $\rm 35^{\circ} \times 35^{\circ} \times \pi$ for each bubble (from \cite{Predehl:2020}), we calculated the total flux of $\rm 6.5 \pm 0.9 \times 10^{-8}~ergs~cm^{-2}~s^{-1}$ and $\rm 4.1\pm 0.5 \times 10^{-8}~ergs~cm^{-2}~s^{-1}$ for the northern and southern bubbles, respectively. Further assuming a distance of $\rm 10.6~kpc$ (from \cite{Predehl:2020}), we estimated their luminosities to be $\rm 8.7\pm1.3 \times 10^{38}~ergs~s^{-1}$ and $\rm 5.6\pm0.8 \times 10^{38}~ergs~s^{-1}$.

\vspace{0.7cm}
\noindent
{\bf Data Availability}: The data presented in this paper are publicly available at the HEASARC (High Energy Astrophysics Science Archive Research Center) archive. 

\vspace{0.7cm}
\noindent
{\bf Acknowledgements}: This research has made use of data obtained from the \suzaku satellite, a collaborative mission between the space agencies of Japan (JAXA) and the USA (NASA). We are grateful to Prof. Barabara Ryden for her notes of the ``Radiate Gas Dynamics'' graduate course at Ohio State. We gratefully acknowledge support through the NASA ADAP grants 80NSSC18K0419 to AG and NNX16AF49G to SM. 

\vspace{0.7cm}
\noindent
{\bf Author Contributions}: A.G. did the X-ray data analysis and wrote the text. S.M. contributed to the interpretation of the results and revised the manuscript. J.K. did the \suzaku data reduction and made the images. S.D. and Y.K. helped with the interpretation of the results. All the coauthors contributed to the discussion and commented on the manuscript.  

\vspace{0.7cm}
\noindent
{\bf Competing Interests}: The authors declare that they have no competing financial interests.

\vspace{0.7cm}
\noindent
{\bf Correspondence and requests for materials}: should be addressed to A.G.

\newpage

%-----------------------------Figure Start------------------------------
\begin{figure}[H]
\renewcommand{\figurename}{Fig. 1}
\renewcommand{\thefigure}{}
\includegraphics[scale=0.6]{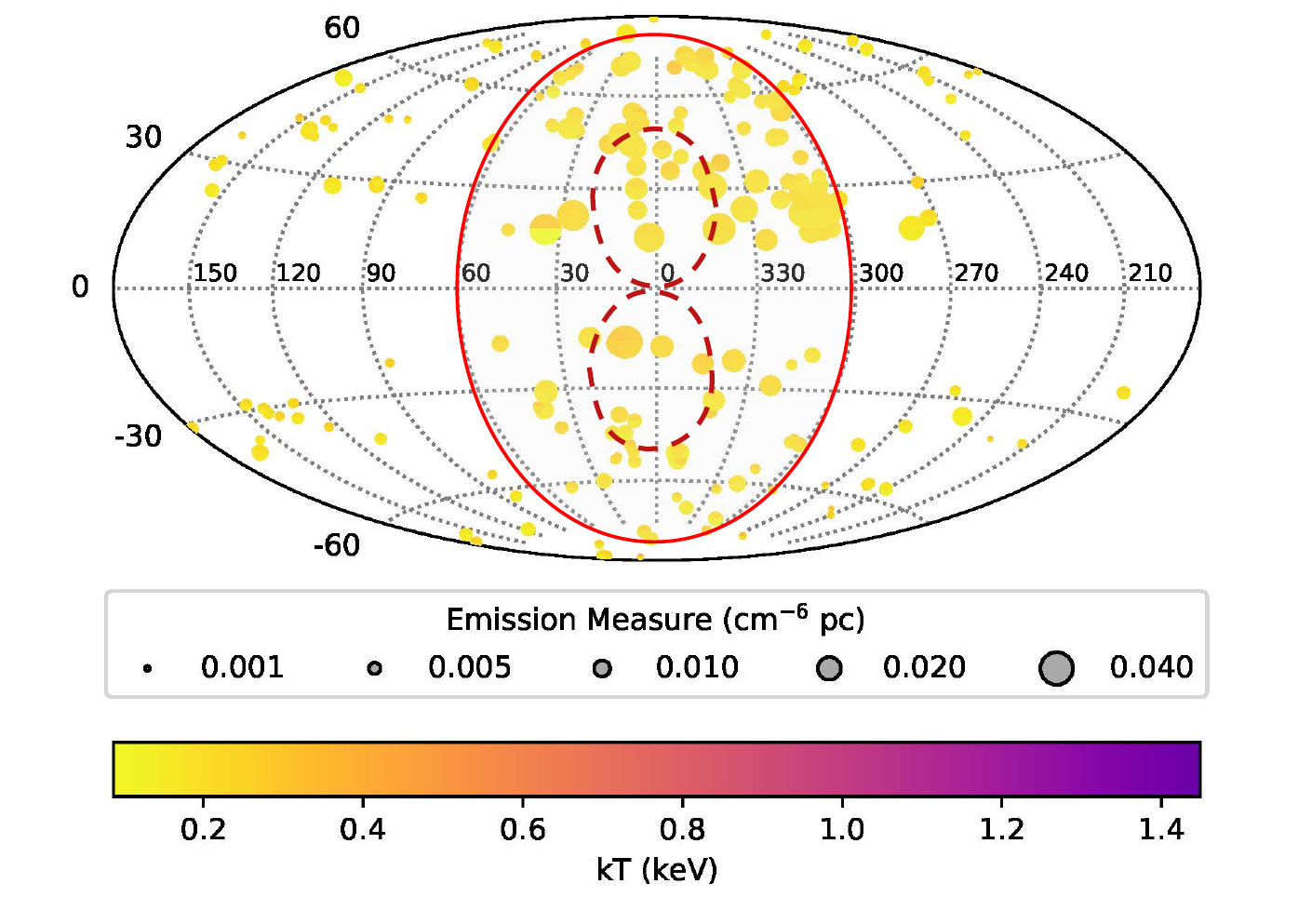}
\\
\includegraphics[scale=0.6]{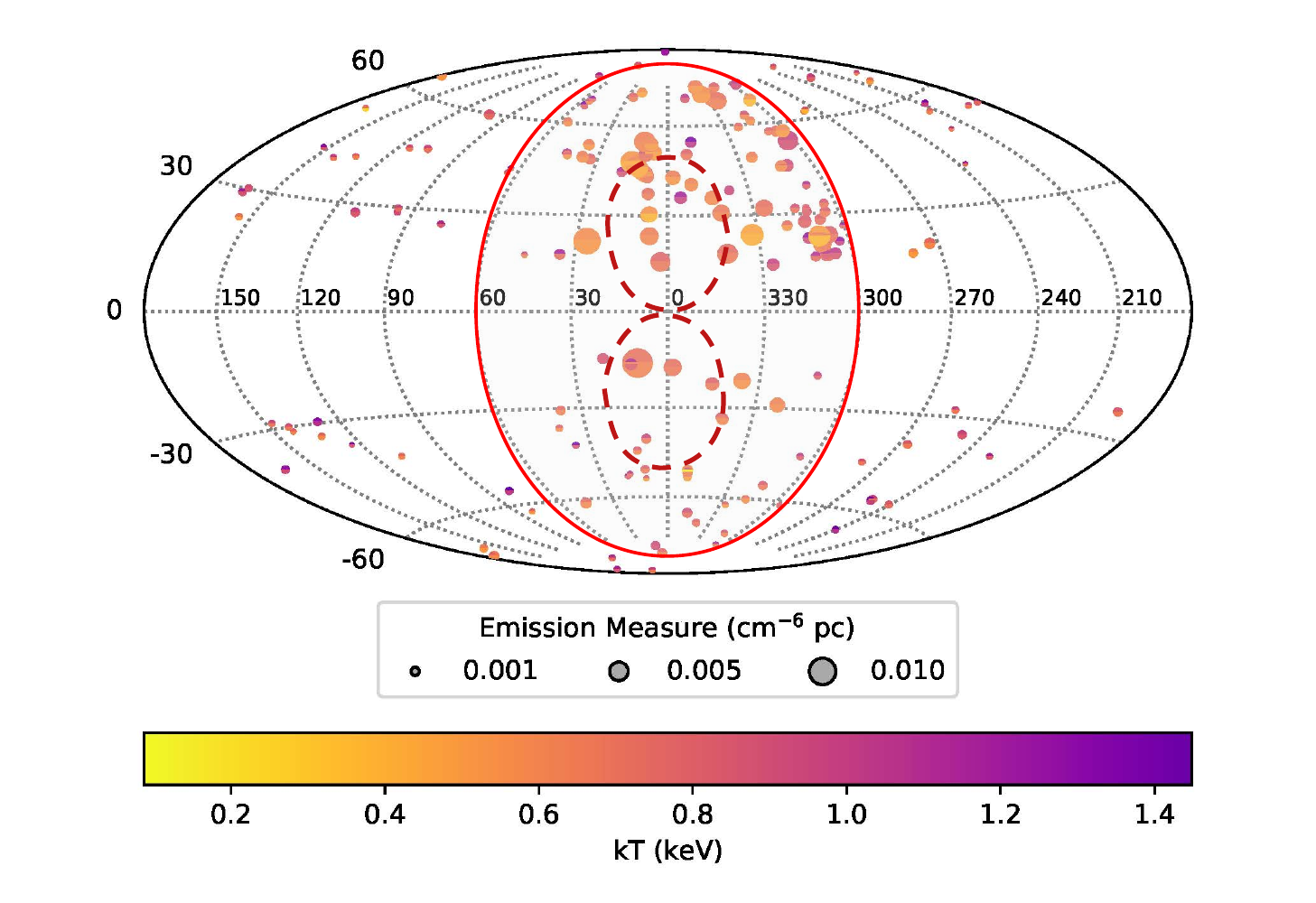}
\vspace*{-0.3 cm}
\caption{X-ray emission maps from our \suzaku survey of the Galactic bubbles and the surrounding halo regions. Figures on {\it top} and {\it bottom} show the distribution of the warm-hot and the hot phases, respectively. The color of each circle indicates temperature while the radius is proportional to the emission measure. The solid red line marks X-ray eROSITA bubbles and the red dashed lines represents the edge of the $\gamma-$ray Fermi bubbles.
%As can be seen, the temperature of the warm-hot component is consistent with Galaxy's virial temperature of $\rm \approx 2 \times 10^{6}~K$ among all sightlines. Whereas, the temperature of second component is much higher ($\rm kT \approx 4-13 \times 10^{6}~K$).
}
%\vspace*{-0.3 cm}
\label{fig1}
\end{figure}
%-----------------------------Figure End--------------------------------

\newpage

%-----------------------------Figure Start------------------------------
\begin{figure}[h]
\renewcommand{\figurename}{Fig. 2}
\renewcommand{\thefigure}{}
\includegraphics[scale=1.0]{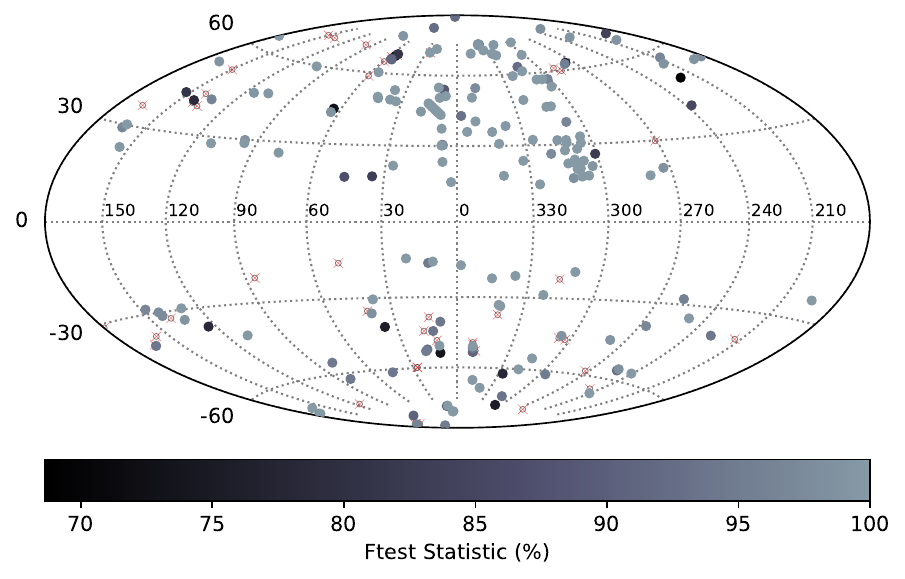}
\vspace*{-0.3 cm}
\caption{F-test probability map for the hot-component significance required over the standard three-component SDXB model for the Suzaku observations investigated in this
work. Empty circles with red crosses mark the sightlines where adding a hot thermal component did not improve the fit.
}
%\vspace*{-0.3 cm}
\label{fig2}
\end{figure}
%-----------------------------Figure End--------------------------------

\newpage
%-----------------------------Figure Start------------------------------
\begin{figure}[h]
\renewcommand{\figurename}{Fig. 3}
\renewcommand{\thefigure}{}
\includegraphics[scale=0.55]{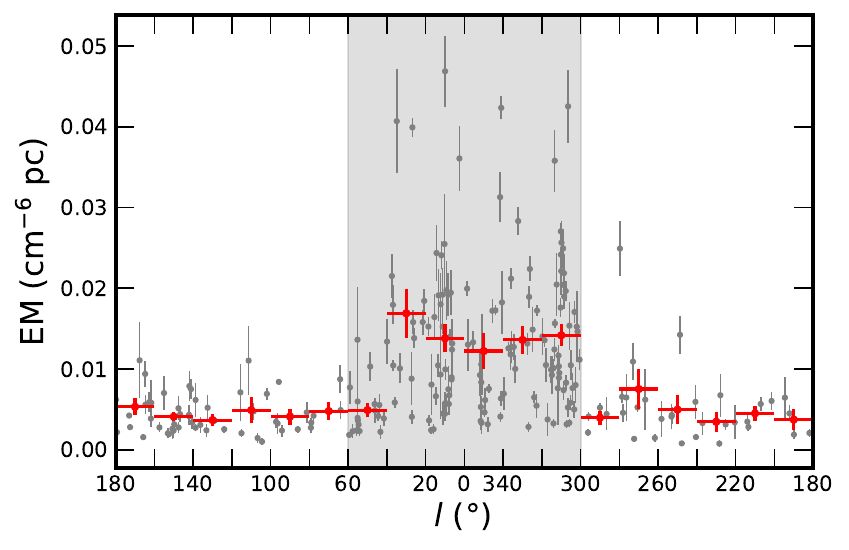}
\includegraphics[scale=0.55]{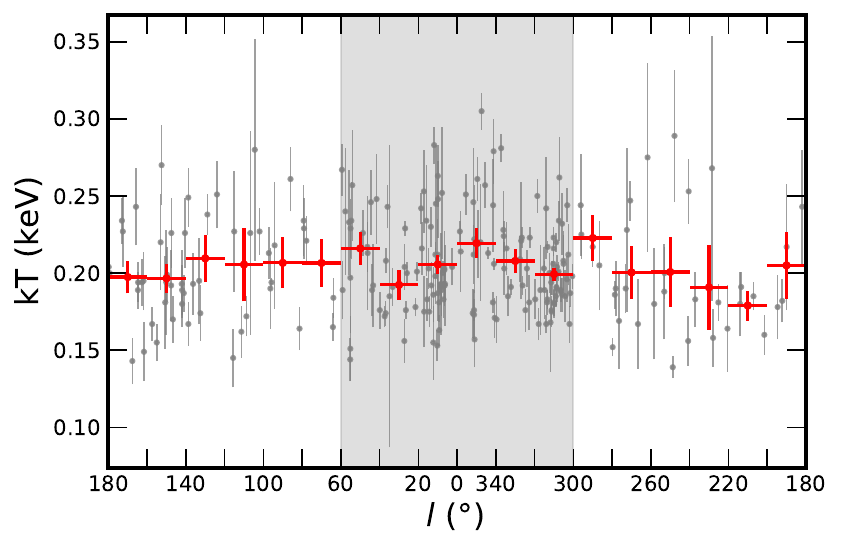}
\\
\includegraphics[scale=0.55]{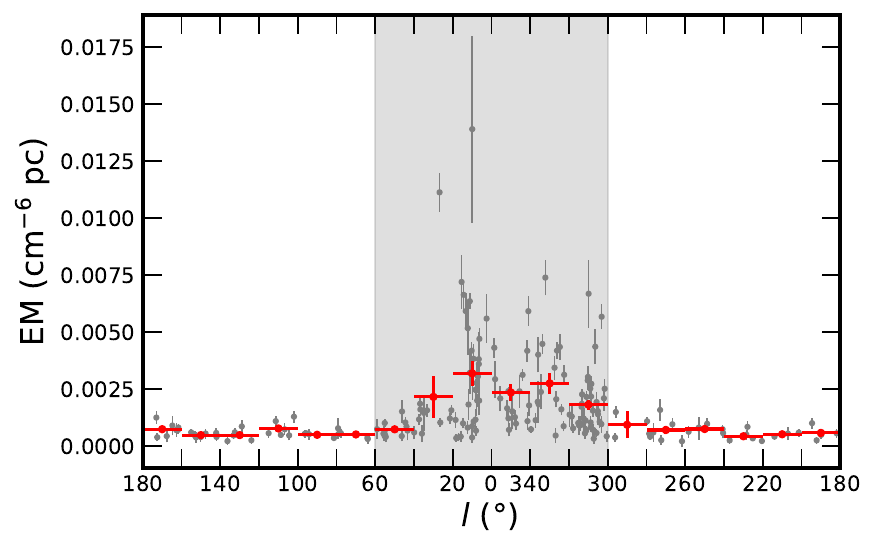}
\includegraphics[scale=0.55]{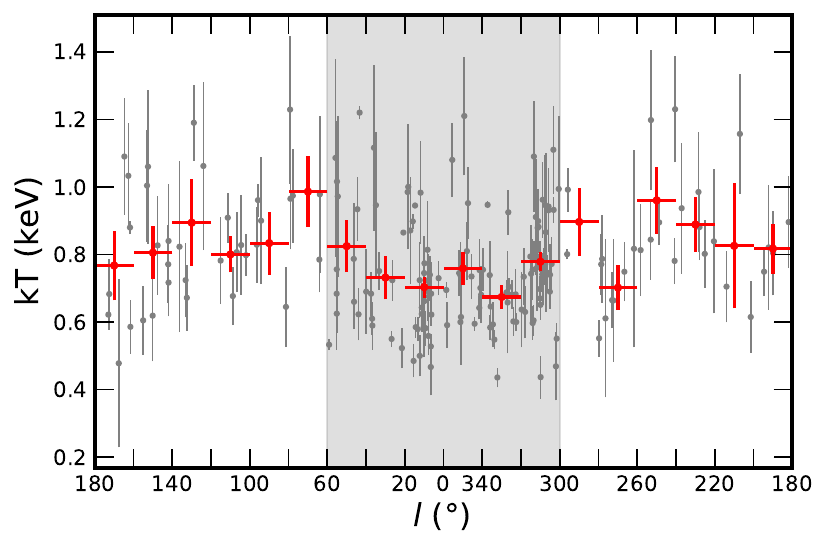}
%\vspace*{-0.3 cm}
\caption{Distribution of the emission measures and the temperatures of the warm-hot (top panels) and the hot (bottom panels) components of the X-ray emission ($\rm b > 15^{\circ}$ \& $b<-15^{\circ}$).  The reported errors are of $1\sigma$ significance. The Galactic bubbles region is shown by the grey shaded band. The red vertical bars include errors as well as the dispersion of the data over $10^{\circ}$ bins.
}
%\vspace*{-0.3 cm}
\label{fig3}
\end{figure}
%-----------------------------Figure End--------------------------------

%\end{document}

\newpage

%-----------------------------Figure Start------------------------------
\begin{figure}[ht]
\renewcommand{\figurename}{Supplementary Fig.~1}
\renewcommand{\thefigure}{}
\includegraphics[scale=0.85]{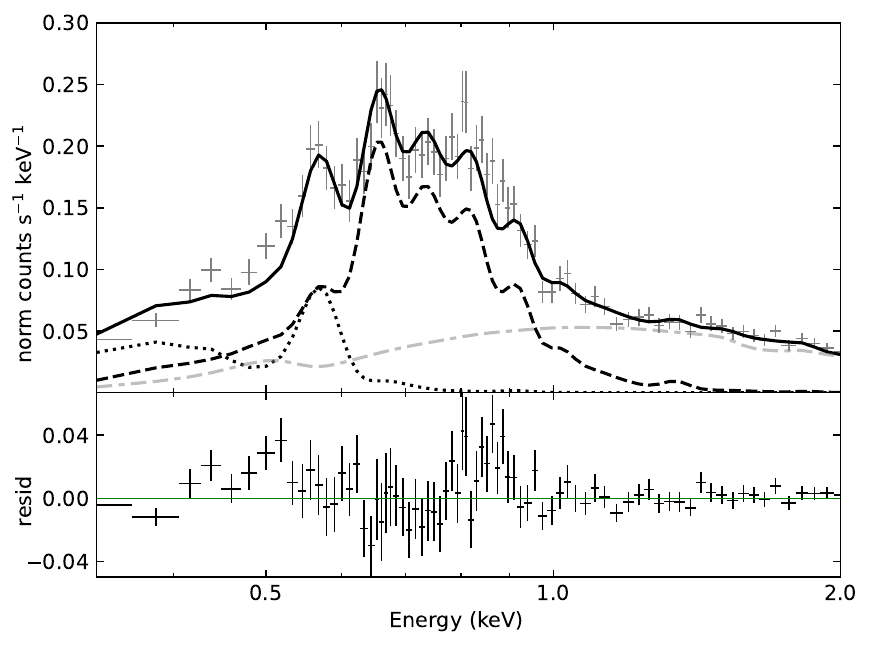}
\vspace{0.5cm}
\includegraphics[scale=0.85]{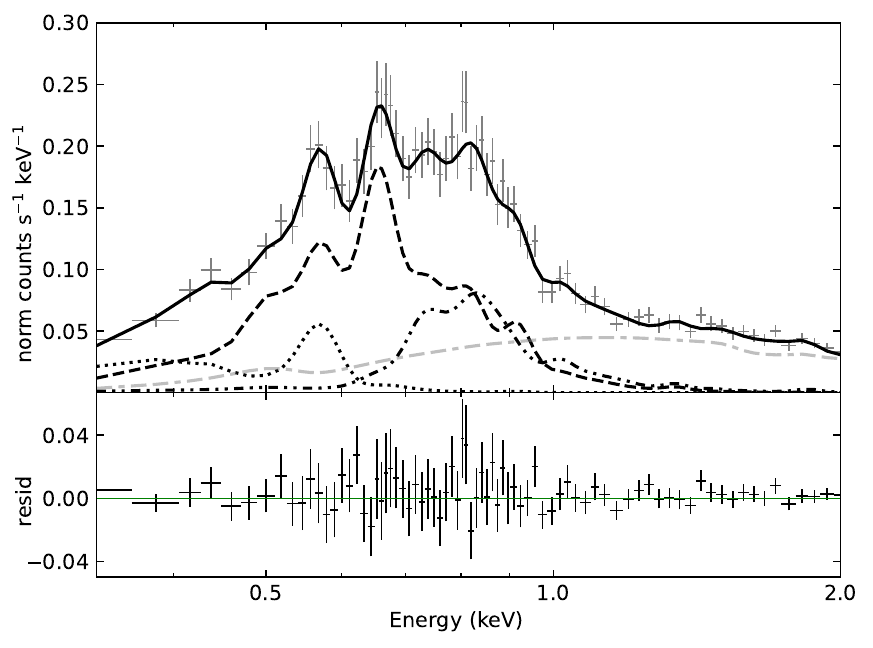}
\vspace*{-0.3 cm}
\caption{\footnotesize Model fits to Suzaku XIS~1 spectra. {\it Top:} One observation investigated in this work with the standard SDXB three component best fit model ($\rm \Delta \chi^{2}/\Delta d.o.f.=374.19/375$). Excess emissions near low (0.4-0.5 keV) and high (0.8-1.0 keV) energy bands can be clearly seen in the residual plot. {\it Bottom:} The best-fit two temperature model with overabundance of nitrogen in the warm-hot phase ($\rm \Delta \chi^{2}/\Delta d.o.f.=349.16/372$). The dotted, dashed and dash-dotted lines indicate the foreground (LHB+SWCX), warm-hot and hot components, respectively. The power-law model of the CXB is shown with the dash-dash-dotted line.
}
%\vspace*{-0.3 cm}
\label{fig4}
\end{figure}
%-----------------------------Figure End--------------------------------

\newpage
%-----------------------------Figure Start------------------------------
\begin{figure}[ht]
\renewcommand{\figurename}{Supplementary Fig. 2}
\renewcommand{\thefigure}{}
\includegraphics[scale=0.55]{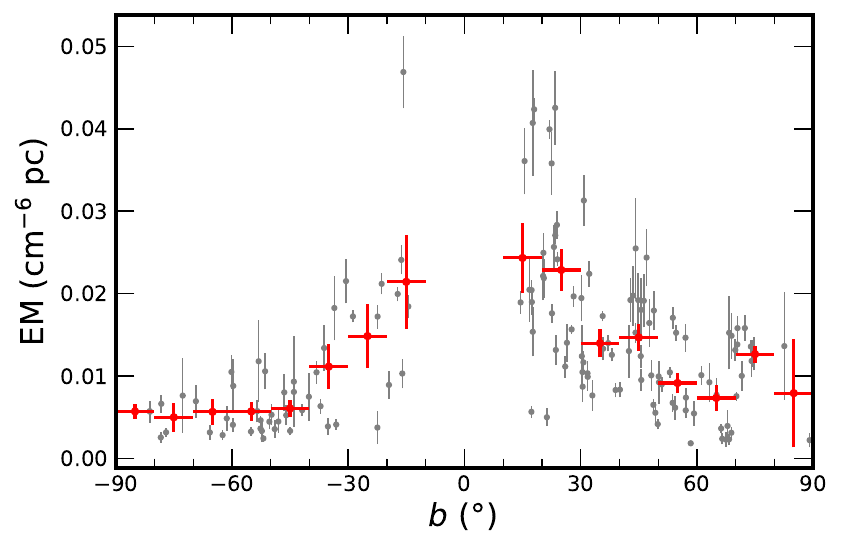}
\includegraphics[scale=0.55]{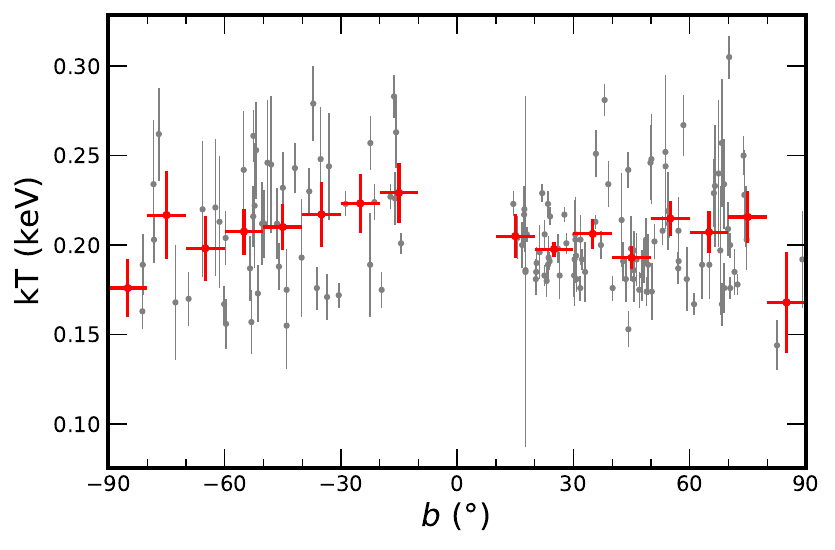}
\\
\includegraphics[scale=0.55]{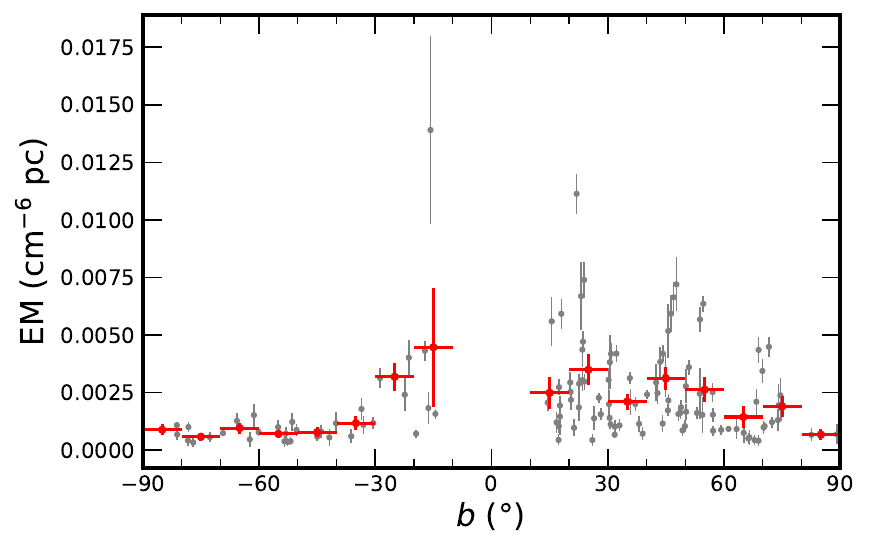}
\includegraphics[scale=0.55]{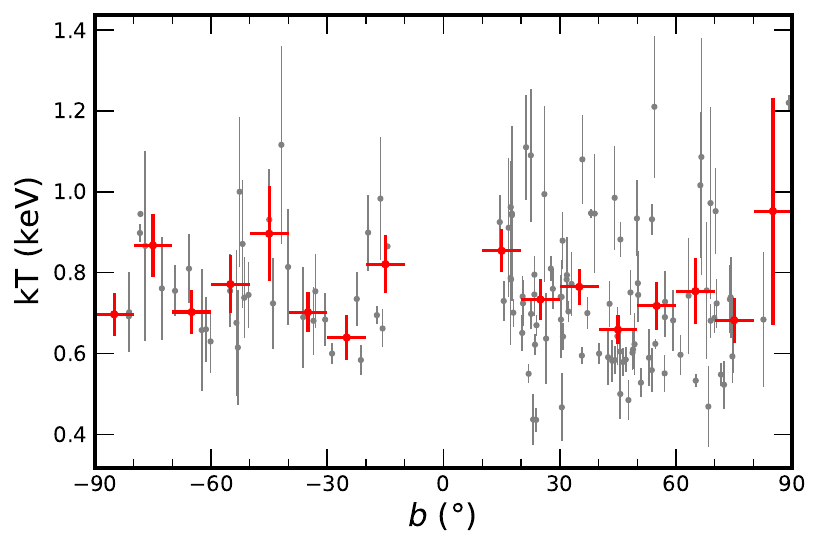}
\vspace*{-0.3 cm}
\caption{Distribution of the emission measures and the temperatures of the warm-hot (top panels) and the hot (bottom panels) in the northern ($\rm b>15^{\circ}$; $\rm 300^{\circ} < l < 60^{\circ}$) and southern bubbles ($\rm b<-15^{\circ}$; $\rm 300^{\circ} < l < 60^{\circ}$). The reported errors are of $1\sigma$ significance.
The red vertical bars include errors as well as the dispersion of the data over $10^{\circ}$ bins.
}
%\vspace*{-0.3 cm}
\label{fig5}
\end{figure}
%-----------------------------Figure End--------------------------------

\newpage
%%%%%%%%%%%%%%%%%%%%%%%%%%%%%%%%%%%%%%%%%%%%%%%%%%%%%%%%%%%%%%%%%%%%%%%%%%

\begin{table}[!h]
\begin{center}
\begin{tabular}{| l l l l l |} 
\hline

{\bf Region} & {\bf $\rm kT_{Mean}$} & &{\bf $\rm EM_{Range}$} & {\bf $\rm EM_{Mean}$}  \\ 
              &  (keV)               & & ($\rm 10^{-3}~cm^{-6}~pc$) & ($\rm 10^{-3}~cm^{-6}~pc$) \\
\hline 
{\bf Warm-Hot Component} & & & &\\
Galactic Bubbles  & $0.205\pm0.004$ & & $2.2-46.9$ & $13.9$\\
Extended Halo & $0.201\pm0.005$ & & $0.8-14.2$ & $4.4$\\ 
 \hline
{\bf Hot Component} & & & &\\
Galactic Bubbles  & $0.741\pm0.018$ & & $0.4-13.9$ & $2.3$\\
Extended Halo & $0.837\pm0.028$ & & $0.2-1.5$ & $0.6$\\ 
\hline
\end{tabular}
\caption{Supplementary Table 1: Distribution of Thermal Parameters of the Warm-Hot and Hot Components.  $1\sigma$ errors are quoted for the temperature. For the EM, we provide the range of the distribution and the mean. }
\end{center}
\end{table}

\newpage

\begin{table}[!h]
\begin{center}
\begin{tabular}{|l l |} 
 \hline
 {\bf Acronyms} & {\bf Definition} \\ 
 \hline 
MW	&	Milky-way\\
GC	&	Galactic Center\\
CGM 	&	Circum-galactic medium\\
SDXB	&	Soft diffuse X-ray background\\
LB	&	Local Bubble\\
SWCX	&	Solar wind charge exchange\\
CXB	&	Cosmic X-ray background\\
EM	&	Emission measure\\
AGN	&	Active Galactic Nuclei\\
BI	&	Back illuminated\\
XIS	&	X-ray imaging spectrometer\\
FI	&	Front illuminated\\
COR	&	cut-of-rigidity\\
ELV	&	Limb of the Earth\\
DYE\_ELV	&	Elevation angle from the bright Earth limb\\
XRT	&	X-ray telescope\\
RMF	&	 Redistribution matrix file\\
ARF	&	Ancillary response file\\
CIE	&	 Collisional ionization equilibrium\\
R-H	&	Rankine–Hugoniot\\
ISM	&	 Interstellar medium\\
 \hline
\end{tabular}
\caption{Supplementary Table 2: List of Acronyms.}
\end{center}
\end{table}

\end{document}